\documentstyle[12pt,amssymb,amscd]{article}
\newcommand{\nc}{\newcommand}
\newcommand{\rnc}{\renewcommand}
\nc{\dbar}{\bar{\partial}}
\nc{\be}{\begin{equation}}
\nc{\ee}{\end{equation}}

\catcode`\@=11

\@addtoreset{equation}{section}
\def\theequation{\thesection\arabic{equation}}

\def\@normalsize{\@setsize\normalsize{15pt}\xiipt\@xiipt
\abovedisplayskip 14pt plus3pt minus3pt%
\belowdisplayskip \abovedisplayskip
\abovedisplayshortskip  \z@ plus3pt%
\belowdisplayshortskip  7pt plus3.5pt minus0pt}
\def\small{\@setsize\small{13.6pt}\xipt\@xipt
\abovedisplayskip 13pt plus3pt minus3pt%
\belowdisplayskip \abovedisplayskip
\abovedisplayshortskip  \z@ plus3pt%
\belowdisplayshortskip  7pt plus3.5pt minus0pt
\def\@listi{\parsep 4.5pt plus 2pt minus 1pt
            \itemsep \parsep
            \topsep 9pt plus 3pt minus 3pt}}

\def\underline#1{\relax\ifmmode\@@underline#1\else
        $\@@underline{\hbox{#1}}$\relax\fi}
\@twosidetrue
\relax

\catcode`@=12

\catcode`\@=11

\def\section{\@startsection{section}{1}{\z@}{3.5ex plus 1ex minus
   .2ex}{2.3ex plus .2ex}{\large\bf}}
\def\thesection{\arabic{section}.}

\def\ps@headings{\def\@oddfoot{}\def\@evenfoot{}
\def\@oddhead{\hbox{}\hfill
        \makebox[.5\textwidth]{\raggedright\ignorespaces --\thepage{}--
        \hfill }}
\def\@evenhead{\@oddhead}
\def\subsectionmark##1{\markboth{##1}{}}
}

\ps@headings

\catcode`\@=12

\relax

\def\figcap{\section*{Figure Captions\markboth
        {FIGURECAPTIONS}{FIGURECAPTIONS}}\list
        {Fig. \arabic{enumi}:\hfill}{\settowidth\labelwidth{Fig. 999:}
        \leftmargin\labelwidth
        \advance\leftmargin\labelsep\usecounter{enumi}}}
 \relax
\def\tablecap{\section*{Table Captions\markboth
        {TABLECAPTIONS}{TABLECAPTIONS}}\list
        {Table \arabic{enumi}:\hfill}{\settowidth\labelwidth{Table 999:}
        \leftmargin\labelwidth
        \advance\leftmargin\labelsep\usecounter{enumi}}}
 \relax
\def\reflist{\section*{References\markboth
        {REFLIST}{REFLIST}}\list
        {[\arabic{enumi}]\hfill}{\settowidth\labelwidth{[999]}
        \leftmargin\labelwidth
        \advance\leftmargin\labelsep\usecounter{enumi}}}
 \relax

\catcode`\@=11

\def\ps@headings{\def\@oddfoot{}\def\@evenfoot{}
\def\@oddhead{\hbox{}\hfill
        \makebox[.5\textwidth]{\raggedright\ignorespaces --\thepage{}--
        \hfill }}
\def\@evenhead{\@oddhead}
\def\subsectionmark##1{\markboth{##1}{}}
}

\ps@headings

\relax

\def\firstpage#1#2#3#4#5#6{
\begin{document}

\begin{titlepage}
\nopagebreak
\title{\begin{flushright}
        \vspace*{-1.8in}
        {\normalsize hep-th/0208081}\\[4mm]
\end{flushright}
\vfill
{\large \bf #3}}
\author{\large #4 \\ #5}
\maketitle
\vskip -7mm
\nopagebreak
\begin{abstract}
{\noindent #6}
\end{abstract}
\vfill
\begin{flushleft}
\rule{16.1cm}{0.2mm}\\[-3mm]
August  2002
\end{flushleft}
\thispagestyle{empty}
\end{titlepage}}
\newcommand{\dal}{\raisebox{0.085cm}
{\fbox{\rule{0cm}{0.07cm}\,}}}
\newcommand{\dt}{\partial_{\langle T\rangle}}
\newcommand{\dtbar}{\partial_{\langle\bar{T}\rangle}}
\newcommand{\al}{\alpha^{\prime}}
\newcommand{\mst}{M_{\scriptscriptstyle \!S}}
\newcommand{\mpl}{M_{\scriptscriptstyle \!P}}
\newcommand{\dv}{\int{\rm d}^4x\sqrt{g}}
\newcommand{\lv}{\left\langle}
\newcommand{\rv}{\right\rangle}
\newcommand{\ph}{\varphi}
\newcommand{\sbar}{\,\bar{\! S}}
\newcommand{\xbar}{\,\bar{\! X}}
\newcommand{\fbar}{\,\bar{\! F}}
\newcommand{\zbar}{\,\bar{\! Z}}
\newcommand{\tbar}{\bar{T}}
\newcommand{\ubar}{\bar{U}}
\newcommand{\ybar}{\bar{Y}}
\newcommand{\phb}{\bar{\varphi}}
\newcommand{\cm}{Commun.\ Math.\ Phys.~}
\newcommand{\pr}{Phys.\ Rev.\ D~}
\newcommand{\pl}{Phys.\ Lett.\ B~}
\newcommand{\prl}{Phys.\ Rev.\ Lett.~ }
\newcommand{\ibar}{\bar{\imath}}
\newcommand{\jbar}{\bar{\jmath}}
\newcommand{\np}{Nucl.\ Phys.\ B~}
\newcommand{\e}{{\rm e}}
\newcommand{\gsi}{\,\raisebox{-0.13cm}{$\stackrel{\textstyle
>}{\textstyle\sim}$}\,}
\newcommand{\lsi}{\,\raisebox{-0.13cm}{$\stackrel{\textstyle
<}{\textstyle\sim}$}\,}
\date{}
\firstpage{95/XX}{3122} {\large\sc Proving the PP-Wave/${\rm
CFT}_2$ Duality} {Edi
Gava$^{a,b}$ and K.S. Narain$^{ b}$ \,}
{\normalsize\sl
$^a$Istituto Nazionale di Fisica Nucleare, sez.\ di Trieste, and SISSA,
Italy\\[-3mm]
\normalsize\sl $^b$The Abdus Salam International
Centre for Theoretical Physics,
I-34100 Trieste, Italy\\[-3mm]}
{We study the duality between IIB string theory on a pp-wave
background, arising as a Penrose limit of the $AdS_3 \times
S^3\times M$, where $M$ is $T^4$ (or $K3$), and the 2D CFT which
is given by the ${\cal N}=(4,4)$ orbifold $(M)^N/S_N$, resolved by
a blowing-up mode. After analizying the action of the supercharges
on both sides, we establish a correspondence between the states of
the two theories. In particular and for the $T^4$ case, we
identify both massive and massless oscillators on the pp-wave,
with certain classes of excited states in the resolved CFT
carrying large $R$-charge $n$. For the former, the excited states
involve fractional modes of the generators of the ${\cal N}=4$
chiral algebra acting on the $Z_n$ ground states. For the latter,
they involve, fractional modes of the $U(1)^4_L\times U(1)^4_R$
super-current algebra acting on the $Z_n$ ground states. By using
conformal perturbation theory we compute the leading order
correction to the conformal dimensions of the first class of
states, due to the presence of the blowing up mode. We find
agreement, to this order, with the corresponding spectrum of
massive oscillators on the pp-wave. We also discuss the issue of
higher order corrections.}

\section{Introduction}
The recent discovery of exactly solvable pp-wave string
backgrounds as particular
double scaling limits of AdS type IIB backgrounds
\cite{blau, metsaev, mt} has made possible to
extend the AdS/CFT correspondence
beyond the supergravity approximation on the string theory side
and therefore, in the dual CFT, beyond the class of strictly
protected chiral primary operators. This has been first achieved
in the context of the $AdS_5\times S^5$ / ${\cal N}=4$ duality\cite{bmn}
and then extended to some of its variations \cite{etal}, including a case
involving open strings/fundamental matter \cite{bgmnn}.
Another interesting dual pair which has been extensively studied
in the past \cite{m,ms,db,mms,agmn} is given by IIB string
theory on $AdS_3\times S^3\times M$,
where $M$ is either $T^4$ or $K3$ on one side and an ${\cal N}=(4,4)$
two-dimensional CFT correponding to the sigma model $M^N/S_N$,
In the Penrose limit this background gives rise an exactly solvable
(in the light-cone gauge, Green-Schwarz formulation)
pp-wave \cite{bmn, rt} string background. Purpose of this paper
is to provide some detailed quantitative  tests of the duality between
string theory on this pp-wave and above 2D CFT.
The first step towards establishing the duality is of course to
provide a correspondence between the states of the two theories:
on the string theory side we have the ground state(s) and the states
built with massive and massless (for $T^4$ case) oscillators, including
zero modes. We will propose a precise correspondence, guided
by the action of supersymmetry on the states on the two theories.
On the CFT side the corresponding class of states is characterized
large $R$-charge $J$, $J \sim \sqrt N$, and $J$ itself is identified
with the $n$ of $Z_n$ twisted sectors.

After identifying ground states with chiral primary
states  of the CFT and massless oscillator modes with
first level descendants of the left-moving ${\cal N}=4$ algebra
generators $L_1$, $G^-_{-\frac{1}{2}}$, $J^-_0$,
(and similarly for the  right-moving  one,
$\tilde {L}_1$, $\tilde {G}^-_{-\frac{1}{2}}$,
$\tilde {J}^-_0$)
we move on to massive oscillators, which will be identified
with certain, say, left fractionally moded descendants,
with generators $L_{-1+\frac{k}{J}}$,
$G^-_{-\frac{1}{2}+\frac{k}{J}}$, $J^-_{-\frac{k}{J}}$,
with $k<< J$, acting on the $Z_J$ ground states.
These states  are
right-chiral at the orbifold point, but become non-chiral
when the orbifold is resolved by switching on
a $Z_2$ blowing up mode.  The necessity of moving off the orbifold
point or, in other words, the fact that the point
in moduli space where the CFT is dual to the pp-wave
is not the orbifold point, is suggested by the
spectrum of the light-cone Hamiltonian, $p^-$,
to be identified in the CFT with $\Delta -J$ , $\Delta$ being the
conformal dimension of the state.
The spectrum is composed of two contributions. Let us
consider  the first one, which is due to the massive
oscillators from the pp-wave part of the background,
and, therefore, has the typical square-root form, implying that
the conformal dimensions of the corresponding states admit
an expansion in powers of $g_6^2 N/J^2$, where $g_6$ is the 6-dimensional
string coupling, much like what happens in the Yang-Mills case \cite{bmn}.
We will see that switching on the blowing-up mode
has the effect of lifting the class of states in consideration,
which are right-chiral (and therefore belong to short
multiplets) at the orbifold point and of making four of them to
join into long
multiplets. As a condequence, their conformal dimensions
are unprotected and are expected to receive corrections.
We will verify this fact by showing that the (right-moving)
supercharges are indeed modified in presence of the blowing
up mode.

The occurrence of this phenomenon will be also supported by a partition
function computation on the two theories, which, as we will see,
also suggests the presence of a supercharge with  a $Z_2$ twist nature,
in agreement with the explicit CFT argument.
We will perform a CFT calculation of the leading order change
in the conformal dimension, finding agreement with the first order
expansion of the square root formula from the pp-wave side.

The second term in the $p^-$ Hamiltonian is due, in the $T^4$
case, to the usual massless oscillators (we do not consider momenta
and windings along $T^4$). This contribution is proportional
to $1/J$. We propose that on the CFT side these correspond
to fractional modes of the $U(1)_L^4\times U(1)^4_R$
super-current algebra acting on the ground states on both
left- and right-moving sectors,
subject to the level matching condition.
First order CFT perturbation theory gives, also for these states,
results in qualitative agreement with the predictions from string theory
spectrum.

The paper is organized as follows. In section 2, we briefly review
$AdS_3/{\rm CFT}_2$ duality and its Penrose limit. The
supersymmetry generators of the pp-wave limit of the  ${\cal
N}(4,4)$ algebra are constructed in section 3. In particular we
find that the anti-commutator of the left and right supercharges
is proportional to the world-sheet momentum operator $P_{\sigma}$
and therefore vanishes on the physical states. In section 4, we
present the computation of the first order corrections to the
conformal dimensions in the boundary CFT. Finally we discuss the
question of higher order corrections and, assuming a natural
identification between the states on the two sides, we show that
the exact expression for the conformal dimensions imply an
extension of the superalgebra: the anticommutator between the left
and right supercharges vanishes on the orbifold group invariant
states but not on individual fractional oscillators. This is
exactly analogous to the results obtained in section 3 for the
pp-wave strings.

While
this paper was in progress, three papers appeared, \cite{hs,lm}
and \cite{gms} addressing some of the
issues discussed here and overlapping partially with ours.

\section{$AdS_3/{\rm CFT}_2$ Duality and its Penrose Limit}
In this section we review the basic facts about the duality
between IIB string theory on $AdS_3\times S^3 \times M$ and the
two dimensional boundary CFT corresponding to the symmetric
product $M^N/S_N$. On the string theory side, the
above background arises as near horizon geometry of a system of
$Q_1$ D1-branes and $Q_5$ parallel D5-branes wrapped on $M$, in
the limit where both $Q_1$ and $Q_5$ are large at fixed $Q_1/Q_5$.
The radius $R$ of both $AdS_3$ and $S^3$ is given by $R^2=\alpha '
g_6 \sqrt N$, with $N=Q_1 Q_5$. $g_6$ is the six dimensional
effective coupling given in terms of the IIB string coupling $g_s$
as $g_6=g_s\sqrt{\frac{Q_5}{Q_1}}$. String theory on such
background is believed to be dual to certain two dimensional
${\cal N}=(4,4)$ SCFT in the Neveu-Schwarz sector. The CFT
corresponds to the orbifold sigma model $M^N/S_N$, possibly
blown-up, depending on the values of spacetime moduli which
survive the near horizon limit.

There have been several tests of this duality, mainly at the level
of spectrum of protected operators \cite{ms,db,mms,agmn}. In
particular, as first done in \cite{db}, the elliptic genus of the
CFT has been compared with the one, suitably defined, of
supergravity. The two have been shown to agree for low energy
excitations \cite{db,mms,agmn}. Notice that, in this comparison,
one is  including also multiparticle supergravity states, but our
discussion here will be focused only on supergravity (string
theory) single particle states. For our later purposes, it will be
useful to reconsider such computations. The elliptic genus
involves states of the form (anything, chiral), but actually it
does not keep track of the $J^3_R$ R-quantum numbers of the
right-moving Ramond ground states, since one sets the conjugate
variable ${\tilde y}$ to ${\bar q}^{-1/2}$ in the character valued
partition function\footnote{The elliptic genus is defined in the
Ramond sector, where one usually sets $\tilde y =1$, but we need
to flow to the NS sector, for the reason explained before.}, ${\rm
tr}_{NS}(-)^F q^{L_0-c/24}y^{J^3_L}$. One may ask then what
happens if we keep arbitrary $\tilde y$ for the chiral states in
the partition function. In general one does not expect the
resulting partition function to be a topological quantity,  and
indeed it was already observed in \cite{agmn} that in this case
there is a discrepancy between supergravity and CFT results. We
want to analyze this point in more detail. From the CFT side, the
partition function for $M^N/S_N$ is best obtained via a generating
function $Z_{CFT}=\sum_N p^NZ(M^N/S_N)$ which can be easily
computed once the multiplicities for the single copy  CFT
$c_{CFT}(m,\ell, \bar m, \tilde\ell )$ are given. Taking into
account only states which correspond to single particle
supergravity states, i.e. the states coming from the twisted
sectors corresponding to single cycles, amounts to consider \be
Z_{CFT}(p,q,{\bar q},y, {\tilde y}) = \sum
c_{CFT}(4mn-n^2-{\ell}^2, 4\bar{m}n-n^2-{\tilde\ell}^2)p^n q^m
y^{\ell} {\bar q}^{\bar m} {\tilde y}^{\tilde\ell}\, \ee where the
multiplicities $c_{CFT}$'s are obtained from the single copy CFT
partition function. The single particle partition function for
states of the form (anything, chiral) on the supergravity side is
computed via the prescription of \cite{db}, once the spectrum of
chiral primaries is known (which happens to coincide with  the
cohomology of $M$), and has an expansion of the form\footnote{The
crucial prescription in this derivation has to do with the notion
of $degree$ formulated in \cite{db}, allowing to discuss finite
$N$ in supergravity, and therefore to introduce the variable
$p$.}: \be Z_{\rm sugra}=\sum_{n,m,\ell,\tilde\ell} c_{\rm
sugra}(n,m,\ell)p^nq^m y^{\ell} ({\tilde y} {\bar
q}^{1/2})^{\tilde\ell} \ee Let us recall that the agreement
between elliptic genera for states of low conformal dimensions
($m\leq N/4$) amounts to the equality $Z_{CFT}(p=1,q,y)=Z_{\rm
sugra}(p=1.q,y)$ and $\frac{\partial}{\partial
p}Z_{CFT}(p,q,y)|_{p=1}= \frac{\partial}{\partial p} Z_{\rm
sugra}(p.q,y)|_{p=1}$, that is, the two elliptic genera differ by
an expression of the form $(p-1)^2 g(p,q,y)$. In the above
discussion we have set  $\tilde y={\bar q}^{-1/2}$, which is the
value corresponding to the elliptic genus flowed in the NS sector.
However using the techniques explained in \cite{db,agmn}, one can
repeat the analysis for the case where one considers right- moving
chiral states with the appropriate power of $\tilde y$ given by
their R-charge. We will not repeat here the calculation, but what
one gets is that the CFT and supergravity partition functions
differ by an expression of the form $(1-p{\bar q}^{1/2} \tilde
y)^2 f(p,q,y, {\bar q}^{1/2}\tilde y)$. This suggests that what
can happen in the CFT is that four short multiplets, coming from
the sectors $n$, (1), $n+1$, (2) and $n+2$, (1) respectively can
join into a long multiplet and drop from the elliptic genus. In
other words, one may expect that, as we move in the CFT moduli
space, the Higgs mechanism can take place and assemble short
multiplets (whose number is a multiple of four) into long ones. We
will explicitly prove in the next section that this is indeed what
happens, and we will identify these states with a class of
(massive) string oscillator states on the pp-wave background.

Let us consider now the Penrose
limit of the $AdS_3\times S^3\times M$
background \cite{bmn,gms}.  This is obtained by
blowing up a region near a null geodesic in $AdS_3\times S^3$.
It involves a scaling limit where $R\rightarrow \infty$ with
$\alpha '$, $g_s$ and  $g_6$ finite. The resulting metric becomes
\be ds^2=-2dx^+dx^- -\mu^2(x^2+y^2)dx^+dx^-+d{\vec x}^2+d{\vec y}^2+
ds^2_M\, \label{pp}\ee where $\vec x$ and $\vec y$ are two-
dimensional vectors, parametrizing a four-dimensional transverse
flat space. There are in addition non trivial vev's for the RR
3-form field strength, $H_{+12}=H_{+34}=\mu$.

The light-cone momenta in this background,$p^+=i\partial_{x^-}$
and $p^-=i\partial_{x^+}$, where $p^-$ is the light-cone
hamiltonian, are identified with the following combinations of
charges in the CFT: \begin{eqnarray} p^-&=&\mu(\Delta-J)\nonumber\\
p^+&=&J/\mu R^2\, \label{charges} \end{eqnarray} where $\Delta$ is
the total conformal dimension in the CFT and $J$ is the $U(1)$
R-symmetry charge $J=J^3_L+J^3_R$, with $J^3_{L,R}$ Cartan
currents in the $SU(2)_L\times SU(2)_R$ R-symmetry algebra of the
${\cal N}=(4,4)$ SCFT. Unitarity in the CFT gives $\Delta-J\geq
0$, corresponding to non-negativity of the light-cone hamiltonian.
The ground states of the latter correspond to chiral primary
states in the CFT. The requirement of considering finite energy
excitations in the limit $R\rightarrow \infty$ (i.e. $N\rightarrow
\infty$), demands to take $J\sim \sqrt N$ with $\Delta -J$ finite.

The spectrum of the light-cone hamiltonian is found to be
\cite{bmn,rt} \be p^-=\sum_n N_n\sqrt{1+(\frac{n}{\mu p^+ \alpha
'})^2}+\frac{L^M_0+{\tilde L}^M_0}{\mu p^+ \alpha '}\,
\label{ham}\ee with the level-matching constraint on the
occupation numbers $N_n$: \be \sum_n N_n=L^M_0-{\tilde L}^M_0 \,
\label{lm} \ee
where $L_0^M$ and ${\tilde L}^M_0$ are left- and
right-moving Virasoro generators for the CFT corresponding to $M$.

With the identifications given above, in particular $\mu p^+
\alpha '= J/g_s Q_5$, we can translate the pp-wave spectrum
(\ref{ham}) into the following statement for the CFT spectrum: \be
\Delta -J=\sum_n N_n\sqrt{1+(\frac{ng_s Q_5}{J})^2}+g_s
Q_5\frac{L^M_0+{\tilde L}^M_0}{J}\, \label{hamcft} \ee

Our main goal will be to reproduce (\ref{hamcft}) directly
from the CFT.

\section{Symmetry Generators}

We now identify the ${\cal N}=(4,4)$ generators on the two sides,
namely string theory on the  pp-wave and two-dimensional CFT.

In the pp-wave background, the mass terms break the light-cone $SO(8)$
symmetry
down to $SO(2)\times SO(2)'\times SO(4)$ where the first and second
$SO(2)$
factors act on the directions coming from $AdS_3$ (say directions
$X^1$ and $X^2$) and $S^3$ (say $X^3$ and $X^4$) respectively while
the
$SO(4)$ acts on the tangent space of $T^4$ (or $K_3$). The generators
of the first and the second $SO(2)$ therefore should be identified
with $L_0-\tilde{L}_0$ and $\tilde{j}_0-j_0$ respectively. The total
dimension $\Delta$ and charge $J$ are $\Delta =L_0+\tilde{L}_0$ and
$J=\tilde{j}_0+j_0$. The fields $X^I, I=1,..,4$ are massive while the
directions along the $T^4$ which we denote by $Y^i$ are massless.
The two $SO(8)$ spinors decompose into $\theta^i_{\alpha a}$ and
$\chi^i_{{\dot{\alpha}} \dot{a}}$ where
$i=1,2$
denotes the worldsheet chirality $\alpha$ and $\dot{\alpha}$ denote
the two
components
of the first $SO(4)$ (which breaks down to $SO(2)\times SO(2)'$ due to
the fermion mass term) spinor and anti-spinor respectively and $a$ and
$\dot{a}$ denote the two components of the spinor and anti-spinor of
the second $SO(4)$ acting on the tangent space of $T^4$. These
fermions satisfy the reality conditions $\bar{\theta}^i_{\alpha a}=
\epsilon^{\alpha \beta} \epsilon^{ab} \theta^i_{\beta b}$ and
similarly $\bar{\chi}^i_{\dot{\alpha} \dot{a}}=
\epsilon^{\dot{\alpha} \dot{\beta}} \epsilon^{\dot{a} \dot{b}}
\chi^i_{\dot{\beta} \dot{b}}$.
While $\theta$ are massive with mass being
the same as that of $X^I$, $\chi$ are massless. The equations of
motion for the massive fields are:
\be
\partial_+ \partial_ X^I = (\mu p^+)^2 X^I, ~~~\partial_+
\theta^1_{\alpha a} -i\mu p^+
(\sigma^3)_{\alpha}^{~\beta}\theta^2_{\beta a}=0, ~~~\partial_-
\theta^2_{\alpha a} +i u p^+
(\sigma^3)_{\alpha}^{~\beta}\theta^1_{\beta a}=0;
\ee

The conserved supercharges have the following expression:
\begin{eqnarray}
Q^c_{\alpha a}&=&\sqrt{p^+}\int d\sigma (e^{i\mu X^+
\sigma_3})_{\alpha}^{\beta}
\theta^c_{\beta a} , ~~~c=1,2 \nonumber\\
Q^{1\dot{\alpha}}_a&=&\int d\sigma (\partial_-
X^{\mu}\bar{\sigma}^{\mu \dot{\alpha} \beta} \theta^1_{\beta a} -i \mu
{p^+}
X^{\mu} \bar{\sigma}^{\mu \dot{\alpha} \beta} (\sigma_3
\theta^2)_{\beta a} + \partial_- Y^i \sigma^i_{a \dot{b}} \chi^{1
\dot{\alpha} \dot{b}}\nonumber\\
Q^{2\dot{\alpha}}_a &=&\int d\sigma (\partial_-
X^{\mu}\bar{\sigma}^{\mu \dot{\alpha} \beta} \theta^2_{\beta a} -i \mu
{p^+}
X^{\mu} \bar{\sigma}^{\mu \dot{\alpha} \beta} (\sigma_3
\theta^1)_{\beta a} + \partial_- Y^i \sigma^i_{a \dot{b}} \chi^{2
\dot{\alpha} \dot{b}}
\end{eqnarray}
$c$ in the first equation above denotes the two world-sheet
chiralities and $(\sigma_3 \theta)_{\alpha a}$ denotes
$(\sigma_3)_{\alpha}^{~\beta}\theta_{\beta a}$.

The mode expansions
for the massless fields $Y^i$ and $\chi$ are standard with the clear
separation of left and right movers. The zero modes of $\chi$ generate
the cohomology of $T^4$: the two complex left-moving zero modes and
two complex right-moving zero modes acting on the ground state
reproduce the cohomology elements $h_{p,q}$ of  $T^4$.

For later convenience it is better to organize the fields in terms of
definite $SO(2)\times SO(2)'$ charges. Thus we define two complex
bosons $Z^1=X^1+iX^2$ and $Z^2=X^3+iX^4$. The complex conjugate fields
will be denoted as $Z^{\bar{i}}$.
the mode expansion for the
massive fields $Z^I$ and $\theta$ are
\begin{eqnarray}
Z^i(\sigma, \tau)&=&
i\sum_{k}\frac{1}{\sqrt{2}\omega_k}(e^{-i\omega_k\tau+ ik\sigma}
a^{i}_k-e^{i\omega_k\tau- ik\sigma} a^{\bar{i} \dagger}_k),
~~~~~i=1,2
\nonumber\\
\theta_{\alpha a}^1(\sigma,\tau)&=&\bigl{[}
\sum_{k\geq 0}c_k e^{-i\omega_k\tau}(e^{ik\sigma}
(b_k)_{\alpha a}+ \frac{\omega_k- k}{\mu p^+}e^{-ik\sigma}
(b_{-k})_{\alpha a}) \bigr{]}+
\nonumber\\ &~&~~~~~\epsilon_{\alpha \beta} \epsilon_{ab}
\sum_{k\geq 0}c_k e^{i\omega_k\tau}(e^{-ik\sigma}
(b_k^{\dagger})^{\beta b}+ \frac{\omega_k- k}{\mu p^+}e^{ik\sigma}
(b_{-k}^{\dagger})^{\beta b})
\nonumber\\
\theta_{\alpha a}^2(\sigma,\tau)&=&-\bigl{[}
\sum_{k\geq 0}c_k e^{-i\omega_k\tau}(e^{ik\sigma}
(\sigma_3 b_k)_{\alpha a}+ \frac{\omega_k- k}{\mu p^+}e^{-ik\sigma}
(\sigma_3 b_{-k})_{\alpha a}) \bigr{]}
\nonumber\\ &~&~~~~~+\epsilon_{\alpha \beta} \epsilon_{ab}
\sum_{k\geq 0}c_k e^{i\omega_k\tau}(e^{-ik\sigma}
(\sigma_3 b_k^{\dagger})^{\beta b}+ \frac{\omega_k- k}{\mu p^+}e^{ik\sigma}
(\sigma_3 b_{-k}^{\dagger})^{\beta b})
\label{mode}
\end{eqnarray}
where
\be
\omega_k=\sqrt{k^2 +(\mu p^+)^2},
\ee
and the normalization constant
\be
c_k= \frac{\mu p^+}{\sqrt{2\omega_k (\omega_k-k)}}
\ee
The oscillator modes satisfy the usual (anti-) commutation relations
\be
[a^i_k, a^{\bar{j} \dagger}_{\ell}]=[a^{\bar{i}}_k, a^{j
\dagger}_{\ell}]=\delta_{ij} \delta_{k\ell},~~~\{ b_{k \alpha a},
b^{\dagger \beta
b}_{\ell} \} = \delta_{\alpha}^{\beta} \delta_a^b \delta_{k \ell}
\ee
Now we are ready to identify the symmetry generators of
${\cal{N}}=(4,4)$ algebra. The left and right supersymmetry generators
are:
\begin{eqnarray}
G_{1/2}^{+ a}&=&(b_0)_{2}^a,~~~~~~  \tilde{G}_{1/2}^{+a} = (b_0)_{1}^a
\nonumber\\
G_{-1/2}^{- a}&=& (b_0^{\dagger})^{2a},~~~~~~
\tilde{G}_{-1/2}^{- a}=(b_0^{\dagger})^{1a}
\nonumber\\
G_{1/2}^{-a}&=& \frac{1}{\sqrt{p^+}}((Q^1)^{\dot{1} a}+(Q^2)^{\dot{1}
a}),~~~~~~~
\tilde{G}_{1/2}^{-a}=\frac{1}{\sqrt{p^+}}
((Q^1)^{\dot{1} a}-(Q^2)^{\dot{1} a}),
\nonumber\\
G_{-1/2}^{+a}&=& \frac{1}{\sqrt{p^+}}((Q^1)^{\dot{2} a}+(Q^2)^{\dot{2}
a}),~~~~~~~
\tilde{G}_{-1/2}^{+a}=\frac{1}{\sqrt{p^+}}((Q^1)^
{\dot{2} a}-(Q^2)^{\dot{2} a}).
\label{GQ}
\end{eqnarray}
The bosonic left and right $SL(2 R)$ and $SU(2)$ generators are:
\begin{eqnarray}
L_1&=& a^2_0,~~~ L_{-1}= a^{\bar{2} \dagger}_0,\nonumber\\
\tilde{L}_1&=& a^{\bar{2}}_0,~~~
\tilde{L}_{-1}= a^{2 \dagger}_0, \nonumber\\
J^+_0&=&a^{\bar{1}}_0,~~~ J^-_0= a^{1 \dagger}_0,\nonumber\\
\tilde{J}^+_0&=&a^{1}_0,~~~ \tilde{J}^-_0= a^{\bar{1} \dagger}_0
\label{LJ}
\end{eqnarray}

For later purposes we give an explicit expression for the part of
$\tilde{G}$ that depends on massive fields:
\begin{eqnarray}
\tilde{G}^{+a}_{-1/2} &=&
\frac{1}{\sqrt{2 p^+}}\sum_{k}[\sqrt{\omega_k+p^+}
(a^{2 \dagger}_k (b_k)_1^{~a} + a^1_k (b^{\dagger}_k)^{1 a})\nonumber\\
&+& {\rm sign}(k) \sqrt{\omega_k-p^+}
(a^{1 \dagger}_k (b_k)_2^{~a} + a^2_k (b^{\dagger}_k)^{2 a})] \nonumber\\
\tilde{G}^{-a}_{1/2} &=& \frac{1}{\sqrt{2 p^+}}\sum_{k}[\sqrt{\omega_k+p^+}
(a^{\bar{1} \dagger}_k (b_k)_1^{~a} -
a^{\bar{2}}_k (b^{\dagger}_k)^{1 a})\nonumber\\
&+& {\rm sign}(k)\sqrt{\omega_k-p^+}
(a^{\bar{2} \dagger}_k (b_k)_2^{~a} - a^{\bar{1}}_k (b^{\dagger}_k)^{2 a})]
\label{G}
\end{eqnarray}
The expressions for $G^{\pm a}_{\mp 1/2}$ is the same as that of
$\tilde{G}^{\mp a}_{\pm 1/2}$ with $\sqrt{\omega_k+p^+}$ exchanged by
$\sqrt{\omega_k-p^+}{\rm sign}(k)$. One can verify that the
anti-commutation relations
between these $G$'s are correct. In particular the anticommutator
$\{G^-_{1/2}, G^+_{-1/2}\}$ defines $L_0-j_0$ and similarly for the
tilde operators. As a consistency check one finds that
\begin{equation}
\{G^{-a}_{1/2}, G^{+b}_{-1/2}\} +
\{\tilde{G}^{-a}_{1/2},
\tilde{G}^{+b}_{-1/2}\}
=\epsilon^{ab}(L_0-j_0 + \tilde{L}_0-\tilde{j}_0) =
\epsilon^{ab}(\Delta - J)
= \epsilon^{ab} P^-
\end{equation}
Note that while $G^{\pm}_{\mp 1/2}$ anticommutes with
$\tilde{G}^{\mp}_{\pm 1/2}$ as they should, with $\tilde{G}^{\pm}_{\mp 1/2}$
they anti-commute only on physical states in the pp-wave string
theory, i.e. the
states that satisfy the level matching condition. Indeed we find
\begin{equation}
\{G^{-a}_{1/2}, \tilde{G}^{-b}_{1/2}\} = \epsilon^{ab} \frac{1}{2p^+}
\sum_k k N_k = \epsilon^{ab} \frac{1}{2p^+} P_{\sigma}
\label{pconj}
\end{equation}
where $N_k$ is the oscillator number operator at momentum $k$ on the
pp wave side and $P_{\sigma}$ just denotes the total momentum. This
might seem strange, but does not create any problem since on the
physical states these anti-commutators vanish. We will see in section
5, the analog of the pp-wave physical state condition as well as the
analog of this anti-commutator in the dual perturbed
boundary CFT.

From this identification it is clear that the insertion of zero mode
creation operators of the massive fields on the pp-wave side
correspond to the insertions of $L_{-1}$, $\tilde{L}_{-1}$, $J^-_0$,
$\tilde{J}^-_0$ (bosonic) and $G^{-\pm}_{-1/2}$,
$\tilde{G}^{-\pm}_{-1/2}$ on the boundary CFT side.

How about the non-zero modes on the pp-wave side. It is natural to
identify them with the insertions of fractional modes of $L_{-1+k/n}$,
$\tilde{L}_{-1+k/n}$, $J^-_{k/n}$, $\tilde{J}^-_{k/n}$ (bosonic) and
$G^{-\pm}_{-1/2+k/n}$, $\tilde{G}^{-\pm}_{-1/2+k/n}$ acting on the
$Z_n$ twisted sector of the boundary CFT (here $k$ can be positive or
negative but is assumed
that $|k| << n$ \footnote{ note that $J^-_{k/n}$ for positive but
small $k$ are still creation operators acting on the $Z_n$ twisted
chiral state which carries a positive charge equal to $(n-1)/2$, $n/2$
or $(n+1)/2$ corresponding to different cohomologies of the $T^4$. The
different cohomology elements are obtained by applying the massless
fermion zero modes on the pp wave side}).

However, one immediately comes up against an apparent contradiction to
this proposal. Consider for instance a state in the boundary CFT of
the form $J^-_{k/n} G^{- b}_{-1/2-k/n} |n> \equiv \Psi_1$, where $|n>$
denotes the chiral-chiral ground state in the $Z_n$ twisted sector
with left and right charges being $(n-1)/2$. This state is non-chiral
in the left moving sector but is chiral on the right sector. It
therefore must be in a (intermediate) BPS multiplet. In other words
$\tilde{G}^{+ a}_{-1/2}$ annihilates this state. On the other hand
according to our proposal this state is to be identified with $a^{1
\dagger}_{-k} (b^{\dagger})^{2 b}_{k} | p^+>$, where $p^+$ is
essentially $n$. It is easy to see, from the explicit expression of
$\tilde{G}^{+ a}_{-1/2}$ given in (\ref{G}), that it does not
annihilate this state. In fact it transforms the state into $d(k)a^{1
\dagger}_{-k} a^{1 \dagger}_{k} | p^+>$ where the coefficient $d(k)=
{\rm sign}(k)\sqrt{\frac{\omega_k-p^+}{2p^+}}$ is an odd function of $k$ and in
the large $p^+$ limit it is $k/p^+$. This state according to our
dictionary is proportional to the state $J^{-}_{k/n'}
J^{-}_{-k/n'}|n'> \equiv \Psi_2$ in the boundary CFT for some $n'$. By
examining the charges we can conclude that $n'=n+1$. In any case this
implies that if the dictionary between boundary CFT and pp-wave string
is correct then such states on boundary CFT side must become
non-chiral via some Higgs mechanism, i.e. several (4 in this case)
intermediate multiplets should join together to give a long
multiplet. The fact that the $Z_n$ twisted sector is mapped to the
$Z_{n+1}$ twisted state suggests that this Higgs mechanism is provided
by the exactly marginal $Z_2$ twist field acquiring a non-vanishing
expectation value.

Another way to see this problem is that the state $\Psi_1$ in the
boundary CFT has $\Delta-J=2$ while the corresponding state on the
pp-wave side has $\Delta-J = 2 \sqrt{1+ k^2/(p^+)^2} \sim 2 +
k^2/(p^+)^2 +..$. Since $\tilde{L}_0-\tilde{j} = \{
\tilde{G}^{-a}_{+1/2} \tilde{G}^{+ b}_{-1/2} \}$, the change in the
dimension can be obtained by computing the change in $\tilde{G}$ under
the marginal deformation. In particular, since the change in the
dimension is not zero, the state cannot be a chiral primary. In the
next section we will compute the
first order change in $\tilde{G}^{+a}_{-1/2}$ due to the marginal
deformation and show that the state $\Psi_1$ indeed is mapped to state
$\Psi_2$ with the coefficient $d(k)$.

Assuming that the identification between the pp-wave string states and
the boundary CFT states is correct, we can then summarize the action
of $\tilde{G}$ on the various left moving excitations, in the large
$J$ limit, as follows:
\begin{eqnarray}
\tilde{G}^{+a}_{-1/2}&:& G^{-b}_{-1/2+k/n}|n> \rightarrow
\frac{k}{\sqrt{2}p^+} \epsilon^{ab}J^-_{k/(n+1)}|n+1>~,\nonumber\\
&&L_{-1+k/n}|n>
\rightarrow \frac{k}{\sqrt{2}p^+} G^{-a}_{-1/2+k/(n+1)}|n+1>~, \nonumber\\
\tilde{G}^{-a}_{1/2}&:& G^{-b}_{-1/2+k/n}|n> \rightarrow
\frac{k}{\sqrt{2}p^+} \epsilon^{ab} L_{-1+k/(n-1)}|n-1>~,\nonumber\\
&&J^-_{k/n}|n>
\rightarrow \frac{k}{\sqrt{2}p^+} G^{-a}_{-1/2+k/(n-1)}|n-1>~.
\label{GQ1}
\end{eqnarray}
where $p^+ = J/g_6 \sim n/g_6$. The action of $\tilde{G}$ on the right
moving fields is the usual in
the large $J$ limit. Note that the third and fourth lines
in the above equation
are just the Hermitian conjugate of the first and second lines.

Another question that comes up with this identification is that, while
the massive non-zero mode creation operators commute among themselves
and thereby ordering of these operators is irrelevant in the
definition of the states, on the boundary CFT side, the corresponding
operators do not commute among themselves. While $L_{-1}$,
$G^{-a}_{-1/2}$ and $J^-_0$ commute among themselves (which correspond
to the zero modes on the pp-wave side), the fractional moded operators
do not commute. Thus it would appear that different orderings of these
operators will give rise to different states. The point, however is
that the difference between different orderings vanishes in the sense
of norm and inner products, in the large $J$ limit. Consider for
example the states $(\frac{1}{n}L_{-1+k/n} J^-{-k/n})|n>$ and
$(\frac{1}{n}J^-{-k/n} L_{-1+k/n})|n>$.  The factor $1/n$ is just
included to have properly normalized states (in the large $J$
limit). The difference between these two states is $\frac{k}{n^2}
J^-_{-2}|n>$, whose inner product with either of the above states is
of order $1/n^2$ and the norm is of order $1/n^3$. Thus in the large
$J$ (or equivalently large $n$) limit all the different orderings can
be identified with a single state.

Finally we turn to the pp-wave string states made up of modes of
massless fields. As mentioned earlier, the two $SO(2)$s acting on
the complex fields $X^2$ and $X^1$ respectively are identified
with $L_0-\tilde{L}_0$ and $\tilde{j}-j$ respectively. This is
also clear from eq.(\ref{LJ}). The bosonic massless fields are
neutral with respect to the two $SO(2)$s and therefore The states
created by them must carry $L_0=\tilde{L}_0$ and $j=\tilde{j}$.
The Fermionic massless field $\chi^{\dot{\alpha} \dot{a}}$ are
anti-chiral with respect to the $SO(4)$ which breaks down to the
product of these two $SO(2)$s and therefore carry opposite quantum
numbers with respect to the two $SO(2)$s.  Thus states created by
these operators must have $L_0-j=\tilde{L}_0-\tilde{j}$.  Since
$L_0-j$ measures how far the state is from being chiral, such
states, on the boundary CFT side, must be equally non-chiral on
both left and right moving sectors. It is natural to conjecture
that the positive and negative momentum $k$ modes of the $T^4$ (or
$K3$) operators on the pp-wave side (i.e. left and right moving
modes on the pp-wave side) should be mapped to states obtained by
applying left and right moving $k/n$ fractional moded $T^4$ (or
$K3$) operators in the $Z_n$ twisted sector on the boundary CFT
side. The creation operators (for simplicity we consider here
$T^4$  or in case of $K3$ its orbifold limit) acting on $Z_n$
twisted chiral-chiral primary state are the left moving fermionic
operators $\psi^{+a}_{-1/2-k/n}$ and $\psi^{-a}_{1/2-k/n}$ (for
$a=\pm$) and the bosonic operators $\alpha^i_{-k/n}$
($i=1,\dots,4$) with $k>0$ and similarly for the right movers. On
the pp wave side the light cone Hamiltonian $\Delta-J$ is
proportional to $k/p^+$. This means that on the boundary CFT side
we must show that even after perturbation the dimension is
proportional to $k/n$. In the following we study the first order
correction to supergenerators and show that to this order it gives
a structure constant proportional to $\sqrt{k/n}$.

\section{Correlators in the resolved $(T^4)^N/S_N$ CFT and first order
correction to the conformal dimensions}

As anticipated in the previous section, we need to consider
correlation functions in the symmetric product CFT, perturbed by the
marginal deformation corresponding to the blowing up mode, which
resolves the $Z_2$ orbifold singularity. For earlier work on the
symmetric product CFT see \cite{dmvv,af,lm1,jmr,lm,dmw}.

Although some of the considerations below apply both to the
$(K3)^N/S_N$ and $(T^4)^N/S_N$ cases, the explicit computations will
be performed for the latter case. The first question we want to
address is the modification of the supersymmetry generators when the
blowing up mode is turned on. We will discuss this change to first
order in the marginal perturbation, following an argument given in
\cite{sen} for the case of the Virasoro generators.

Recall that the blowing up mode is obtained from the chiral primary
field $\sigma_{\frac{1}{2},\frac{1}{2}}$ in the $Z_2$ twisted sector,
of left- right- dimensions (1/2,1/2), by applying the left- and
right-moving supercharges $G^{- a}_{-\frac{1}{2}}$ and ${\tilde
G}^{- a}_{-\frac{1}{2}}$, where $a=+,-$ denotes the doublet components
of a global $SU(2)_I$ which is an outer automorphism of the ${\cal
N}=(4,4)$ algebra. Together with the conjugate charges
$G^{+ a}_{-\frac{1}{2}}$, ${\tilde G}^{+ a}_{-\frac{1}{2}}$ they are
part of the global ${\cal N}=(4,4)$ symmetry of the CFT. Here the
explicit $+$ and $-$ refer to the components of a doublet with respect
to the global left- and right $SU(2)_R\times \tilde{SU(2)}_R$
R-symmetry charges, which are also part of the ${\cal N}=(4,4)$
algebra. Therefore there are actually four blowing up modes, which are
top components of a short (4,4) multiplet and are neutral under the
$SU(2)_R$, $\tilde{SU(2)}_R$ Cartan generators, but transform as ${\bf
3}+{\bf 1}$ of $SU(2)_I$. We will be interested in perturbing the
symmetric product CFT with the singlet component, corresponding to a
particular combination of RR zero- and four-form vev's in the type IIB
background.  This is given by an antisymmetric combination of left and
right supercharges acting on the chiral primary field, so that the
complete expression for the perturbation is: \be
\lambda(G^{- +}_{-\frac{1}{2}}{\tilde G}^{- -}_{-\frac{1}{2}}-
G^{- -}_{-\frac{1}{2}}{\tilde G}^{- +}_{-\frac{1}{2}} )
\sigma_{\frac{1}{2},\frac{1}{2}} + {\rm a.c.} ~~~~~~,
\label{pert}
\ee where $\lambda$ is the coupling constant, a.c. refers to the
expression involving the antichiral field
$\bar{\sigma}_{\frac{1}{2},\frac{1}{2}}$ with the hermitean conjugate
supercurrents $G^{+ \pm}$, ${\tilde G}^{+ \pm}$.

We will be interested in studying the action of the right-moving
charges ${\tilde G}^{+ a}_{-\frac{1}{2}}$, on states (or fields) which
are chiral primary on the right-moving sector and excited on the
left-moving sector. By definition, ${\tilde G}^{+ a}_{-\frac{1}{2}}$
annihilates such states if the blowing up modes are switched off. So
let us consider bringing down in a correlator involving these fields
the blowing up mode of (\ref{pert}) to first order, and look at the
action of ${\tilde G}^{+ a}_{-\frac{1}{2}}$: this is represented by a
contour integral $\oint_{C} d\bar{z} {\tilde G}^{+ a}(\bar{z})$ where
the contour $C$ encircles the positions of all the other fields,
including the perturbation which is integrated over the plane: \be
<\oint_{C} d\bar{z} {\tilde G}^{+ a}(\bar{z})\prod_i
\phi_i(z_i,\bar{z}_i) \int d^2x \sigma_{1,1}(x, \bar{x})>
\label{pert1} \ee where $\sigma_{1,1}=\epsilon_{a b}
G^{- a}_{-\frac{1}{2}}{\tilde G}^{- b}_{-\frac{1}{2}}
\sigma_{\frac{1}{2},\frac{1}{2}}$
is the blowing up mode.
Deforming the contour around $z_i$'s gives zero if the $\phi_i$
are right-chiral. The integral around $x$ can be evaluated using
o.p.e.'s: the interesting term comes from writing \be \sigma_{1,1}(x,
\bar {x})= \oint_{C_x} d\bar{y} \epsilon_{a b}{\tilde
G}^{- a}(\bar{y})\sigma^b_{1,\frac{1}{2}}(x,\bar {x})~,
\label{desc}
\ee and using the o.p.e. of ${\tilde G}^{+ \pm}(\bar z)$ with ${\tilde
G}^{- \mp}(\bar w)$. In (\ref{desc}) $\sigma^b_{1,\frac{1}{2}}=
G^{- b}_{-\frac{1}{2}}\sigma_{\frac{1}{2},\frac{1}{2}}$.

The simple pole in this o.p.e is given by:

\be{\tilde G}^{+ \pm}(\bar z) {\tilde G}^{- \mp}(\bar w)|_{\rm simple~
pole}= (\tilde{T}(\bar w)+\bar{\partial}\tilde {J}_3(\bar{w})/(\bar
{z}-\bar{w}), \label{ope} \ee ,

where $\tilde T$ and $\tilde{J}_3$ are respectively the right-moving
stress energy tensor and Cartan current of the $SU(2)_R$ R-symmetry.
Recalling the o.p.e. of $\tilde T$ with $\sigma^a_{1,\frac{1}{2}}(x,\bar
{x})$, one sees that this gives a total derivative in $\bar x$: \be
\int d^2 x {\bar\partial}\sigma^a_{1,\frac{1}{2}}(x,\bar {x})\prod_i
\phi_i(z_i,\bar{z}_i)=\sum_i\oint_{C_{z_i}}dx \sigma_{1,\frac{1}{2}}
(x,\bar {x}) \phi_i(z_i)\prod_{j\neq i} \phi_j(z_j,\bar{z}_j)
\label{contact} \ee This motivates the following expression for the
first order change in the supercharge in the presence of the
blowing up mode:
\be \delta
{\tilde G}^{+ a}_{-\frac{1}{2}}\phi _i(0)=\lambda \oint dx
\sigma^a_{1,\frac{1}{2}}(x,\bar {x})\phi _i(0) \label{deltag} \ee We
have considered here the chiral part of the perturbation, the first
term in (\ref{pert}), however it is easy to see that the antichiral
gives the same result (\ref{deltag}): the reason is that $\sigma$ and
$\bar\sigma$ are members of an $SU(2)_R$ doublet, \be
\bar\sigma_{\frac{1}{2},\frac{1}{2}} =\oint J^-(z) \oint d\bar{z}
{\tilde J}^-(\bar{z})\sigma_{\frac{1}{2},\frac{1}{2}} .
\label{anti}
\ee But then if we move $J^-$ and ${\tilde J}^-$ across $G^+$ and
$\tilde{G}^+$ respectively, we get back $G^-$ and $\tilde{G}^-$, so
that, in fact, the two terms in (\ref{pert}) are identical
\footnote{We have omitted here, and it will also be understood
in the following when not strictly necessary,
the $SU(2)_I$ index on the various fields.}.

We se therefore that there is a change in ${\tilde G}^+$ if there is a simple
pole $1/x$ in the o.p.e between $\phi$ and
$\sigma_{1,\frac{1}{2}}$. If this is the case, and this will be
verified shortly, four states which belong to intermediate multiplets of
the (4,4) supersymmetry at the orbifold point should join into long
multiplets when turning on the blowing up mode.

Moreover, the action of $\delta {\tilde G}^+_{-\frac{1}{2}}$ on a given field
$\phi$ amounts to fusing $\sigma_{1,\frac{1}{2}}$ with $\phi$.  Now
$\sigma_{1,\frac{1}{2}}$ is in the $Z_2$ twisted sector and $\phi$ is
in the $Z_n$ twisted sector, so this gives a state in the $Z_{n+1}$
twisted sector when the corresponding two-cycle $(i _1,i_2)$ and
$n$-cycle $(j_1,\dots j_n)$ have one index in common.  There will be
also a transition from $Z_n$ to $Z_{n-1}$ when the two-cycle is
contained in the $n$-cycle, and this will be the essentially CPT
conjugate amplitude involving $\delta {\tilde G}^-_{\frac{1}{2}}$ with the
exchange of in- and out-states
\footnote{Notice that the proper twist fields should correspond to
conjugacy classes of $S_N$, and those corresponding to $n$-cycles can
be obtained from the twist field corresponding to a given $n$-cycle by
averaging over the $S_N$ group orbit.}.
Following basically the same steps leading to (\ref{deltag}),
i.e. inserting $\delta {\tilde G}^{-a}_{\frac{1}{2}}$  in the correlator
of (\ref{pert1}) as $\oint_C \bar{z} {\tilde G}^{-a}(\bar{z})$, deforming the
contour and using the o.p.e. between ${\tilde G}^{- \pm}(\bar z)$ and
 ${\tilde
G}^{+\mp}(\bar w)$, whose relevant part is given by:

\be{\tilde G}^{-\pm}(\bar z) {\tilde G}^{+\mp}(\bar w)\sim
\frac{\tilde{T}(\bar w)-\bar{\partial}\tilde {J}_3(\bar{w}}
{\bar
{z}-\bar{w}} -\frac{2\tilde {J}_3(\bar{w})}{(\bar
{z}-\bar{w})^2}+\cdots, \label{ope} \ee
one can prove the following equation:
\be \delta
{\tilde G}^{-a}_{\frac{1}{2}}\phi _i(0)=\lambda \oint dx
\bar{x}
\bar{\sigma}^a_{1,\frac{1}{2}}(x,\bar {x})\phi _i(0) \label{deltag} \ee
In this case therefore $\delta
{\tilde G}^{-a}_{\frac{1}{2}}$ is determined by the $1/x\bar x$ singularity
in the o.p.e. between $\bar{\sigma}_{1,\frac{1}{2}}(x, \bar{x})$  and
$\phi$.

There is however a priori a possibility of transitions from a
$Z_{n}$-sector state to a $Z_{n-1}$ state which is genuinely
different from the previous one. Notice that in the previous
discussion we had, on the right-moving sector, chiral states which
correspond to a cohomology element with 0 degree on the
right\footnote{The single particle cohomology elements of
$(T^4)^N/S_N$ from a given $Z_n$ twisted sector inherit the
Dolbeault degrees $(p,q)$ of the original $T^4$, with $p,q=0,1,2$.
In the CFT, $p=0$ ($q=0$) corresponds to the basic $Z_n$ twist
field on the left (right), whereas $p=1,2$ ($q=1,2$) correspond to
applying one or two left- (right-) moving invariant fermions,
respectively, with positive $R$-charge. This increses the
dimension and the charge by the same amount so that the resulting
field is still chiral primary.}  Now, $R$-charge conservation
alone on the right-moving sector would allow a transition via a
$Z_2$ twist field from a $Z_n$ chomology element at degree 0 to a
chomology element in the $Z_{n-1}$ of degree 2 (two fermions
applied). Similar considerations hold for the conjugate process.
We will verify in the following, by an explicit computation that
the corresponding amplitude in fact vanishes.

Finally, if the incoming two- and n-cycles are disjoint, then one
obtains an outgoing cycle which corresponds to a two-particle state
and this will not be of interest for us.

The above picture agrees with the result found in section 2, where
we have seen that the discrepancy between the supergravity and CFT
single-particle partition functions, for states which are right-moving
chiral, is of the form $(1-\bar{q}^{\frac{1}{2}}\tilde{y}p)^2
f(p,q,\bar{q}^{\frac{1}{2}}\tilde{y})$, with some given function $f$.
This indeed means that the three supermultiplet's components come from
$(n,n+1,n+2)$ twisted sectors, in agreement with the $Z_2$ twisted
character of the modified supercharges discussed above.

We would like now to analyze in more detail the fusion of
$\sigma_{1,\frac{1}{2}}$ with the left-moving excited field in the
$Z_n$-twisted sector. We will be interested in a particular class
of such fields, namely those obtained by applying an arbitrary
string of fractionally moded generators $L_{-1+\frac{k}{n}}$,
$G^{-,a}_{-\frac{1}{2}+\frac{l}{n}}$ and $J^{-}_{\frac{m}{n}}$,
where $L$'s and $J^-$'s are the modes of the stress energy tensor
and of the negatively charged R-symmetry current respectively. The
integers $k,l,m$ are assumed to be much smaller than $n$ in
absolute value, i.e.  $|k|,|l|,|m|<< n$ and have to satisfy
orbifold group invariance  condition, which demands, in the
present case, that the total fractional mode should vanish, i.e.
$\sum l_i+\sum k_i +\sum m_i =0$. This condition corresponds
precisely to the level matching condition on the pp-wave side,
equation (\ref{lm}).

Before discussing the relevant three-point functions, let us
consider first the unperturbed two-point functions involving one
of these fields and its conjugate. This correlator will involve
multiple insertions of an even number of $T$'s, of pairs of
$G^+$'s and $G^-$'s and of $J^+$'s and $J^-$'s. The particular
mode insertion can be then obtained by a contour integration
around the positions of the chiral and conjugate anti-chiral
external fields. The following observation will be useful when we
will discuss the perturbed two point functions: in the large $n$
limit we are interested in, the leading contribution to this
correlator comes from "diagonal" contractions, where each of the
above generators is contracted with a conjugate generator. The
reason for this comes from the fact that in the o.p.e. of two
conjugate operators one finds schematically $T$, $T+\partial J_3$
or $J_3$. When bringing these operators near the chiral primary in
the $Z_n$ twisted sector, they will give a contribution of order
$n$, because they will measure the dimension or the charge (or the
sum of the two) of the corresponding field, which is $(n-1)/2$.
Off-diagonal contractions will be suppressed with respect to the
diagonal ones by powers of $n$. This argument also says that in
order to have two-point functions of order 1, we should normalize
the generators with a $1/\sqrt{n}$ factor.

This observation facilitates the analysis of the three point
function, i.e. the two point function involving the blowing up
mode to first order, as far as the leading contribution for large
$n$ is concerned. In this case of course the presence of the
blowing up mode forces the two external excited states to be
non-diagonal. However, from the previous argument, the leading
contribution will arise when all the generators inserted will be
contracted diagonally, apart from a pair which has to connect
through $\sigma_{1,\frac{1}{2}}$. The upshot of this argument is
that one can concentrate on the part of the correlator which
involves a "flip" of the generators via the interaction
$\sigma_{1,\frac{1}{2}}$. Charge conservation then restricts the
consideration to two type of correlators\footnote{The following
correlators are understood as part of a correlator which involve
states satisfying the orbifold group invariance condition.} :
\begin{eqnarray} C_1(u) &=&<n+1|J^+_{-\frac{k}{n+1}}
\sigma_{1,\frac{1}{2}}(u,\bar u)
G^-_{-\frac{1}{2}+\frac{k}{n}}|n>,\nonumber\\ C_2(u)
&=&<n+1|G^+_{+\frac{1}{2}-\frac{k}{n+1}} \sigma_{1,\frac{1}{2}}(u,
\bar u) L_{-1+\frac{k}n{}}|n>, \label{3point}
\end{eqnarray}

where $|n>$ and $|n+1>$ denote ground states in the $Z_n$ and
$Z_{n+1}$ twisted sector respectively.

Before turning to a more detailed computation, let us observe that the
$u$, $\bar u$ dependence of the three-point functions in
(\ref{3point}) (or rather of the full three-point function, involving
physical states obeying level-matching condition) is fixed by the
conformal dimensions of the three fields, and therefore it is of the
form $c_{n,n+1}/u$, where $c_{n,n+1}$ is a structure constant, and we
have assumed that the external twist fields are at $0$ ($Z_n$) and
$\infty$ ($Z_{n+1}$). Therefore the non-trivial task for our purposes
is the determination of the leading $n$ dependence of the structure
constants, and this will give us the action of the modified
right-moving supercharges on the excited states we are interested in.

In order to perform the computation of the above correlators it is
convenient to introduce a covering map from the plane $t$, on which
the correlator is single valued, the original plane $z$ on which it
has the appropriate monodromies. This map can be chosen to be: \be
z=t^n(t-t_0) ,\label{map} \ee with some arbitrary parameter $t_0$.
This is a branched covering, with branch point of order $n$ at $t=0$,
of order $n+1$ at $t=\infty$ and of order 2 at $x=nt_0/(n+1)$, where
the $Z_2$ twist fields is located. This map thus realizes the required
monodromies. Denoting with $u$ the value of $z$ at $t=x$, we have \be
u=z(x)=-x^{n+1}/n. \label{phys} \ee In terms of the $t$ coordinate the
scalar correlator is simply \be {\partial} X(z){\partial} {\bar X}(w)=
\frac{t'(z)t'(w)}{(t(z)-t(w))^2 } \label{scalar}\ee It is also
convenient to bosonize the two complex fermions $\Psi^{+\pm}_I$ by
introducing two canonically normalized scalar fields $H_1^I$ and
$H_2^I$ as follows \begin{eqnarray}
\Psi^{++}_I&=&\exp{(iH_1^I+iH_2^I)/\sqrt{2}}\nonumber\\
\Psi^{+-}_I&=&\exp{(iH_1^I-iH_2^I)/\sqrt{2}}, \label{fermion}
\end{eqnarray} which both carry charge $1/2$ with respect to
$J_3=(\sum_I H^I_1)/\sqrt{2}$. Obvious expressions, with signs
flipped, hold for the conjugate fields $\Psi^{--}$ and $\Psi^{-+}$.
The index $I=1,\dots, N$ refers to the $I$-th field copy. As for the
twist field in the $Z_n$ twisted sector, it carries charge $(n-1)/2$
with respect to $J_3$ and therefore its $H_1$ dependence is given by
\be \exp(i\frac{(n-1)\sum H_1^I}{\sqrt{2}n})~, \label{twist}\ee

whereas it does not depend on $H_2$. The total dimension of the twist
field is given by the sum of the dimension of the charged component,
$(n-1)^2/4n$, and the dimension of the neutral one, $(n^2 -1)/4n$,
which gives a total of $(n-1)/2$, in agreement with its chiral primary
nature.  Similar expressions hold for the right-moving fields.

Let us consider first the three point function $C_1$. It is convenient
to perform the calculation on the covering $t$-plane, where it is easy
to find the following expression:
\begin{eqnarray} C_1=\frac{1}{(n+1)x^{n-1}}\oint_{\infty}dt_2\oint_0 dt_1
&&(t_2)^{-\frac{kn}{n+1}}(t_2-t_0)^{-\frac{k}{n+1}} (t_1)^k
(t_1-t_0)^{\frac{k}{n}}\nonumber\\
&&(\frac{t_2}{t_1})^{n-1}\frac{1}{(t_2-t_1)(t_1-x)^3},
\label{corr} \end{eqnarray} where the contour integrations with
the corresponding $k$-dependent powers project on the given fractional
modes. The latter $t_2$ and $t_1$ dependent factors are, in terms of
the $z$ coordinate, just $z_2^{-\frac{k}{n+1}}$ and
$z_1^{\frac{k}{n}}$ respectively. It is convenient to rescale $t_1$
and $t_2$ by $x$, after which we will get, for $n$ large, an overall
factor $1/x^{n+1}$ in (\ref{corr}), together with the replacement of
$t_0$ by $n/(n+1)$, which is equal to 1 in the large $n$ limit.  The
prefactor $1/x^{n+1}$ was to be expected on dimensional grounds, since
it gives a simple pole in the physical coordinate $u$, in view of the
relation (\ref{phys}), which gives also an additional factor $1/n$ in
the correlator. It remains to evaluate the contour integrals in
(\ref{corr}), which is most easily done by doing a series expansion of
the integrand. More details are given in Appendix A, but is easy to
see that the leading contribution is of order $kn$ for $n$ large. To
estimate the overall behavior of the complete correlator, including
"spectator" generators, one should take into account that each
generator comes with a normalization factor $1/\sqrt n$, as already
mentioned, so that the contribution of the diagonal contractions is of
order 1. Furthermore, by doing the relevant combinatorics, one can
show that the three point function of normalized ground states
$<n+1|\sigma_{\frac{1}{2},\frac{1}{2}}|n>$ goes like $n$ for large
$n$. So, collecting all the factors of $n$, we see that the full three
point function behaves like $k/n$ for large $n$.

The above result could have been obtained perhaps more transparently
by using a superconformal Ward identity which relates $C_1$ to the
correlator \be C'_1=<n+1|J^+_{-\frac{k}{n+1}}
\sigma_{\frac{1}{2},\frac{1}{2}}(u,\bar u) J^-_{-\frac{k}{n}}|n>.
\label{wi} \ee The Ward identity is obtained by rewriting
$G^-_{-\frac{1}{2}+\frac{k}{n}}|n>$ in $C_1$ as $G^+_{-\frac{1}{2}}$
applied to $J^-_\frac{k}{n}|n>$ and doing the contour deformation we
used previously. The only contribution arises when
$G^+_{-\frac{1}{2}}$ hits $\sigma_{1,\frac{1}{2}}(u,\bar u)$, which
gives $L_{-1}\sigma_{\frac{1}{2},\frac{1}{2}}(u,\bar u)$. So we get as
a result of these manipulations that $C_1(u)=\partial_u C'_1(u)$. On
the other hand, on dimensional grounds, the $u$ dependence of $C'_1$
is $u^{\frac{k}{n(n+1)}}$, so that the $u$ derivative produces a
suppression factor $1/n^2$. The $n$ dependence of $C'_1$ itself is
quickly estimated in the following way: to the leading order in $1/n$
the fractional modes of $J^+$ and $J^-$ are opposite to each other, so
that when we bring say $J^+$ against $J^-|n>$ we get $J^3_0 |n>$, that
is a factor proportional to $n$. Taking into account the normalization
of the currents, we see that $C'_1$ is of order 1. Therefore again we
see that $C_1$ is indeed of order $1/n$ in agreement with the previous
analysis, if we take into account the normalization of the three twist
field correlator which goes like $n$.

Let us come now to the other basic correlator $C_2$: again using SCFT
Ward identities we can relate it to $C_1$ plus another correlator.
This is done by writing \be L_{-1+\frac{k}{n}}|n> =(G^+_{-\frac{1}{2}}
G^-_{-\frac{1}{2}+\frac{k}{n}}+\frac{k}{n}J^3_{-1+\frac{k}{n}})|n>~.
\label{c0}
\ee The first term in (\ref{c0}) can be shown, after bringing $G^+$
against the other operators, to be equal to: \be C_1 +
<n+1|G^+_{\frac{1}{2}-\frac{k}{n+1}}
\partial_u\sigma_{\frac{1}{2},\frac{1}{2}}(u,\bar u)
G^-_{-\frac{1}{2}+\frac{k}{n}}|n>~,
\label{c2}
\ee where $C_1$ arises when $G^+_{-\frac{1}{2}}$ acts on
$<n+1|G^+_{\frac{1}{2}-\frac{k}{n+1}}$ and the second term when we
commute $G^+$ with $\sigma_{1,\frac{1}{2}}$.

The term involving $J^3$ in (\ref{c0}) can be shown explicitly,
using the procedure explained in Appendix A, to be suppressed by a
power of $n$ with respect to $C_1$.  We are left therefore with
the second term in (\ref{c2}).  It is easy to see that this term
is suppressed too in the large $n$ limit compared to $C_1$: the
reason is that all the $n$ countings go in the same way as for the
$J^+$, $J^-$ correlator of the previous paragraph, but there is an
additional $1/n$ suppression factor coming from the jacobians
appearing when relating the $t$ coordinates to the $z$
coordinates. They are there in this case (as well as in
(\ref{corr}), giving the prefactor $1/(n+1)$) because the
supercurrents which are being contour integrated have weight
$3/2$, whereas they are absent in the case of the two currents
$J^+$ and $J^-$, which have weight 1, and therefore their contour
integral is invariant under coordinate changes.  Morew details are
given in appendix A.

Let us discuss now amplitudes which involve a change in the
cohomology's degree, and show, as anticipated before, that they do not
occur in the present situation.  To this purpose it is sufficient to
consider the right moving sector of the amplitude: \be <n+1|{\tilde
G}^{-}_{\frac{1}{2}}\sigma_{1,1} \tilde{\Psi}^{++}_{-\frac{1}{2}}
\tilde{\Psi}^{+-}_{-\frac{1}{2}}|n>~,
\label{cd}
\ee One can rewrite this correlator in the following way: \be
<n+1|\oint_\infty dz_3 z_3 {\tilde G}^{-}(z_3)\oint_u dz {\tilde
G}^-(z) \sigma_{1,\frac{1}{2}}(u) \oint_0
\frac{dz_1}{z_1}\tilde{\Psi}^{++}(z_1)
\oint_0\frac{dz_2}{z_2}\tilde{\Psi}^{+-}(z_2)|n>~.
\label{cd1}
\ee This expression is computed using the $t$ coordinates, using the
map $z_i= t_i^n(t_i-t_0)$ and keeping track of the variuos jacobians
as explained already. One can perform then first the $t$, $t_1$ and
$t_2$ contour integrations, but then one finds that the final $t_3$
contour integral vanishes, being of the form $\oint_\infty dt_3
(t_3-t_0)/(t_3-x)^3$.

We can now use the above results to compute, to the first non-trivial
order, the change in the right-moving conformal dimensions of the
left-excited states under discussion.  The starting point is the
asnti-commutation relation:
\be\{\tilde{G}_{+\frac{1}{2}}^-,\tilde{G}_{-\frac{1}{2}}^+\}=
\tilde{L}_0-\tilde{J}^3_0,
\label{alg}
\ee Here ${\tilde G}^{\pm}_{\pm\frac{1}{2}}$ is actually ${\tilde
G}^{(0)\pm}_{\pm\frac{1}{2}}+\delta {\tilde G}^\pm_{\pm\frac{1}{2}}$,
with  $\tilde{G}^{(0)\pm}$ the zeroth order supercharge.

We can take now the matrix element of (\ref{alg}) between any of
the (normalized) excited states proportional
$J^-_{-\frac{k}{n}}|n>$, $G^-_{-\frac{1}{2}+\frac{k}{n}}
|n>$ or $L_{-1+\frac{k}n{}}|n>$ and its conjugate. Notice that these states
are related to each other by supersymmetry.
Consider for example $|\psi_n> \sim J^-_{-\frac{k}{n}}|n>$. From
equation (\ref{3point})  we have:
\begin{eqnarray}
\delta\tilde{L}_0-\tilde{j}=<\psi_n|\delta {\tilde
G}^+_{-\frac{1}{2}} |\psi'_{n-1}><\psi'_{n-1}|\delta {\tilde
G}^-_{+\frac{1}{2}}|\psi_n>
= {\lambda}^2|c_{n-1,n}|^2
\label{deltah}
\end{eqnarray}
Here we have used the fact that the zeroth order ${\tilde
G}^{(0)}+_{-\frac{1}{2}}$ annihilates $|\psi_n>$, so that there is no need
to consider the second order change in ${\tilde G}^+_{-\frac{1}{2}}$.
The excited state $|\psi'_{n-1}>$, from the $Z_{n-1}$ twisted sector,
is proportional to $G^-_{-\frac{1}{2}+\frac{k}{n-1}}|n-1>$ .
From (\ref{deltah}), we see that indeed the leading
order change in the right-moving conformal dimensions for the above
states is of order $(k/n)^2$ for a given mode $k$. What remains
to do is to fix its $N$ dependence. Recall that the
external states, from
the $Z_n$ twisted sector, are canonically normalized, however
we have to fix the normalization of the blowing up mode. If we assume
that the coupling constant $\lambda$ equals the effective 6-dimensional
string coupling $g_6=g_s \sqrt{\frac{Q_5}{Q_1}}$, where $g_s$
is the  string coupling\footnote{Independent arguments
supporting this identification
have been given in \cite{gms}.}, and the $Z_2$ chiral
primary $\sigma$ is actually obtained
by summing over the two-cycles (i,j), i.e. it is of the form
$\sigma=\sum_{i<j}^N\sigma_{ij}$, then the three point function above
is of order $\sqrt N=\sqrt{Q_1 Q_5}$, since the canonically normalized
$Z_2$ twist field is $\sigma/N$ and the three-point function of
canonically normalized twist fields goes like $1/\sqrt{N}$.
Collecting all the factors we get a correction of the form

\be
\frac{1}{2}(g_s)^2 (Q_5)^2 (\frac {k}{n})^2
\ee
for $\Delta= L_0+\bar L_0$, which is in agreement
with the first order expansion of the
pp-wave hamiltonian for massive oscillators, when translated into
CFT units.

We now discuss another class of states, which should correspond to
$T^4$ massless oscillator states in the light-cone hamiltonian on the
pp-wave.  These are non-chiral on both left- and right-moving sectors,
and are obtained by applying fractional modes of the $U(1)_L^4\times
U(1)_R^4$ super-current algebra, generated by the four free bosons and
fermions, to the ground state in the $Z_n$ twisted sector.
Level-matching requires that $L_0-{\tilde L}_0=0\, {\rm mod}\, n$.
Neither of the unperturbed supercharges ${G}^+_{-\frac{1}{2}}$ and
${\tilde G}^+_{-\frac{1}{2}}$ annihilates these states, and to zeroth
order in the blowing up mode both $\tilde{L}_0-\tilde{J}^3_0$ and
${L}_0-{J}^3_0$ are of the form $k/n$, with $k<<n$. The question is
what happens when we switch on the blowing up mode. In general their
conformal dimensions will change, but in order for the change to agree
with the pp-wave string spectrum it should happen that, to any order
in in the blowing up mode coupling constant, the change should be
proportional to $k/n$. In this case the effect of turning on the
blowing up mode will be to renormalize $k/n$ by some finite amount
depending on its coupling constant.  We will not be able to prove
this, but we will perform structure constant calculations similar to
those described previously, which will confirm the above picture,
giving indeed structure constants of order $\sqrt{k/n}$.

Before turning to the calculation of the relevant
three point functions, let us fix the
correct normalization of the fractionally moded bosonic and fermionic
oscillators. We have, using the covering coordinates $t$'s, for the
left-moving bosons:

\be <n|{\bar\alpha}_{\frac{k}{n}}\alpha_{-\frac{k}{n}}|n>= \oint_0
dt_1\oint_\infty dt_2 ({\frac{t_2}{t_1}})^k \frac {1}{(t_1-t_2)^2}=k
~,
\label{bn}
\ee whereas for the left-moving fermions: \be
<n|{\Psi}^{+\mp}_{\frac{1}{2}-\frac{k}{n}}\Psi^{-\pm}_
{-\frac{1}{2}+\frac{k}{n}}|n> =\oint_0
\frac{dt_1}{\sqrt{t_1'}}\oint_\infty \frac{dt_2}{\sqrt{t_2'}}\frac{1}
{(t_1-t_2)}\frac{{t_2}^{n-1}}{t_1^n} ({\frac{t_2}{t_1}})^k= n ~ ,
\label{fn}
\ee Similar expressions hold for the right-movers.

Thus we should rescale each bosonic (fermionic) oscillator by $1/\sqrt
k$ ($1/\sqrt n$) when constructing normalized excited states.

The arguments of the first part of this section concerning the
modification of $G^+$, ${\tilde G}^+$ in the presence of the blowing
up mode go through also for this class of states, the only difference
being that now there is a non-trivial zeroth-order action of $G^+$,
${\tilde G}^+$ as well. In particular, at first order in the
perturbation, we should compute three-point functions involving
external excited states and an insertion of $\sigma_{1,\frac{1}{2}}$
or $\sigma_{\frac{1}{2},1}$.  Let us consider the first case. In order
for the three point function to be non-vanishing, the two external
states should contain operators acting on $Z_n$ and $Z_{n+1}$ ground
states respectively , which, in the right-moving sector, are conjugate
of each other whereas in the left-moving sector they are conjugate
apart from a single flip due to the presence of
$\sigma_{1,\frac{1}{2}}$. $R$-charge conservation then requires in the
left-moving sector there should be a flip from a bosonic oscillator to
a fermionic oscillator carrying $+1/2$ $R$-charge.  The basic
correlators we need to consider are therefore\footnote{We omit here
the $\pm$ index corresponding to the direction 2 in (\ref{fermion})}:
\begin{eqnarray}
D_1(u)&=&<n+1|\Psi^+_{\frac{1}{2}+\frac{k}{n+1}}
\tilde{\alpha}_{\frac{k}{n+1}} \sigma_{1,\frac{1}{2}}(u,\bar u)
{\alpha}_{-\frac{k}{n}}
\tilde{\bar\alpha}_{-\frac{k}{n}}|n>~,\nonumber\\
D_2(u)&=&<n+1|\Psi^+_{\frac{1}{2}+\frac{k}{n+1}}
\tilde{\Psi}^+_{\frac{1}{2}+\frac{k}{n+1}} \sigma_{1,\frac{1}{2}}
(u,\bar u){\alpha}_{-\frac{k}{n}} \tilde{\Psi}^-_
{-\frac{1}{2}-\frac{k}{n}}|n>~,
\label{corr2}
\end{eqnarray}
These correlators can be computed along the lines we followed to
evaluate $C_{1,2}$. Let su first connsider $D_1$: one starts from the
correlator: \be <n+1|\Psi^+(t_2)\bar{\partial} X({\bar t}_2)
\sigma_{1,\frac{1}{2}}(u,{\bar u}) {\partial} X(t_1){\bar \partial}
{\bar X}({\bar t}_1)|n>~,
\label{D1}
\ee and then projects onto the given modes by contour integrating with
the factors $z_1^{\frac{k}{n}}$ and $z_2^{-\frac{k}{n+1}}$ as before.
It is easy to see that the resulting expression , when expressed in
terms of the physical coordinate $u$ goes like $k^2/nu$. Taking into
account the normalization factors discussed above, giving a factor
$1/k^{\frac{3}{2}}\sqrt n$ and the three-ground state correlator,
which goes like $n$, we see that the resulting structure constant
$d^1_{n,n+1}$ in $D_1\sim d^1_{n,n+1}/u$ goes like
$\sqrt{\frac{k}{n}}$.

One can similarly evaluate $d^2_{n,n+1}$ in $D_2$, starting with \be
<n+1|\Psi^+(t_2)\bar{\partial} {\tilde \Psi}^+({\bar t}_2)
\sigma_{1,\frac{1}{2}}(u,\bar u) {\partial} X(t_1) {\Psi}^-({\bar
t}_1)|n>~,
\label{D2}
\ee and proceeds in the same way. In this case the correlator goes
like $ k/u$, so that taking into account the normalization factors,
which give now $1/n^{\frac{3}{2}}\sqrt k$, and usual factor of $n$
from the ground states correlator, gives for the structure constant
$d^2_{n,n+1}\sim \sqrt{\frac{k}{n}}$.

In this case, unlike in the previous case, the first order structure
constants above do not allow to compute the full change in the
conformal dimensions, since the zeroth order supercharges do not
annihilate the states and we need also the second order
expression. So, what is required is really a full fledged four point
function calculation. Nevertheless we believe that the results above
give a good indication that indeed the variations of the conformal
dimensions at arbitrary orders in the blowing up mode are proportional
to $k/n$.

\section{Discussion}

In the last section we saw that the first order correction to the
dimension matches with the prediction from the pp-wave side. Now we
discuss the higher order corrections for the states corresponding to
the massive oscillators.

The relevant set of states are the excitations of the
$Z_n$ twisted chiral-chiral state by applying fractional moded
$J^-_{k/n}$, $G^-_{-1/2+k/n}$ and $L_{-1+k/n}$ operators.
 Let us start from the state $|\psi_n> \equiv N_n J^-_{-\frac{k}{n}}
|n>$ where $N_n$ is the normalization constant so that $|\psi_n>$
has unit norm. This is not a physical state; there must be
multiple insertions of fractional modes so that the total level is
an integer (or half integer for fermionic states). However we have
seen in the first order analysis that in the large $n$ limit, it
is sufficient to consider the interaction of only one of the
insertions, the remaining ones are spectators and contract
diagonally. $(j,\tilde{j})$ charge of $|\psi_n>$ is
$(\frac{n-3}{2},\frac{n-1}{2})$. Using methods given in the last
section one can see that $\tilde{G}^+_{-1/2}$ and $G^-_{1/2}$
annihilates this state at least to the first order in the
perturbation. Moreover $\tilde{G}^-_{1/2} |\psi_n>$ is
proportional to  $G^+_{-1/2} |\psi_{n-1}>$. It seems natural to
assume that to all orders in the perturbation, under the action of
${\cal N}=(4,4)$ generators, the set of states under consideration
are mapped to each other (in large $n$ limit). If this is so then
the above first order results must be true to all orders. This is
seen simply by taking into account the left and right $U(1)$
charges of the states.

Now using the supersymmetry
algebra
we find that the norm square of the state $\tilde{G}^-_{1/2} |\psi_n>$ is
\begin{equation}
||\tilde{G}^-_{1/2} |\psi_n>||^2 = \gamma <\psi_n|\psi_n>
\end{equation}
where $\gamma= \tilde{L}_0-\tilde{j}$ eigenvalue of $|\psi_n>$.

On the other hand, we have seen from the three point computation
given in the last section, that to the first order in
perturbation:
\begin{equation}
<\psi_{n-1}|{G}^{-a}_{1/2}\tilde{G}^{-b}_{1/2} |\psi_n> =
\epsilon^{ab} \frac{\lambda k\sqrt{N}}{2n} \label{conj}
\end{equation}
If we assume that this equation is true to all orders in
$\lambda/n$, then, using the fact that the norm square  of the
state $G^+_{-1/2} |\psi_{n-1}>$ is $L_0- j$ eigenvalue of
$|\psi_{n-1}>$ which is equal to $\gamma+1$ (note that the
correction to $L_0$ and $\tilde{L}_0$ must be equal), we conclude
that:
\begin{equation}
\gamma = \frac{1}{\gamma+1} \frac{\lambda^2 k^2N}{4n^2}
\end{equation}
The solution to this equation is
\begin{equation}
\gamma = \frac{1}{2}(\sqrt{1+\frac{\lambda^2 k^2N}{n^2}} -1)
\end{equation}
Identifying $n/\lambda \sqrt N$ with $p^+$ and the fact that the
correction to $\Delta$ is twice the correction to $\tilde{L}_0$
and hence $2\gamma$, we find complete agreement with the pp-wave
spectrum.

This incidentally also proves the formula for $\tilde{G}$ given in
(\ref{GQ1}). Indeed it follows that
\begin{equation}
\tilde{G}^-_{1/2} |\psi_n> = \frac{\lambda k \sqrt N}{n}
\frac{1}{\sqrt{\gamma+1}} |\psi'_{n-1}>
\end{equation}
where $|\psi'_{n-1}> =\frac{1}{\sqrt{\gamma+1}}
G^+_{-1/2}|\psi_{n-1}>$ has unit norm. Using the value of $\gamma$
and the identification of $p^+$ as above, we find that:
\begin{equation}
\tilde{G}^-_{1/2} |\psi_n> = \sqrt{\frac{\omega_k-p^+}{p^+}} |\psi'_{n-1}>
\end{equation}
which is exactly the relevant term appearing in the expression of
$\tilde{G}$ from the pp-wave side in (\ref{GQ1}) .

The exact correspondence with the pp-wave therefore rests on the
validity of equation (\ref{conj}) to all orders. Since
${G}^-_{1/2}$ annihilates $|\psi_n>$, this is equivalent to the
statement that
\begin{equation}
\{\tilde{G}^{-a}_{1/2}{G}^{-b}_{1/2}\} |\psi_n> = \epsilon^{ab} \lambda
\int d z \partial_z (z\bar{z}  \bar{\sigma}_{\frac{1}{2},\frac{1}{2}})
|\psi_n>
\end{equation}
where the integration contour encircles the insertion of the state
$|\psi_n>$. This equation is the analog of equation (\ref{pconj})
appearing in the pp-wave string, provided the operator on the
right-hand side is identified with $P_{\sigma}/p^+$. The latter
vanishes on the physical states though not on the individual
oscillators. This is also the case with the operator $\int d z
\partial_z (z\bar{z} \bar{\sigma}_{\frac{1}{2},\frac{1}{2}})$
appearing above: physical states consisting of multiple insertions
satisfying $\sum k_i=0$ have integer (or half-integer for
fermionic states) unperturbed dimensions and therefore are local
with respect to $\bar{\sigma}_{\frac{1}{2},\frac{1}{2}}$ and thus
are annihilated by this operator. This is not the case with the
individual building blocks of such states. Indeed the individual
operator insertion carries fractional dimension $k/n$, while the
outgoing state carries dimension $k/(n-1)$. The difference of the
dimensions being of order $k/n^2$, gives rise to the OPE:
\begin{equation}
(z \bar{z} \bar{\sigma}_{\frac{1}{2},\frac{1}{2}}(z,\bar{z})
|\psi_n> \sim n.\frac{k}{n^2} \log z |\psi_{n-1}> \,
\end{equation}
resulting in equation (\ref{conj}). Here the first factor of $n$
on the right hand side comes because there are $n$ channels by
which $\bar{\sigma}$ can take $Z_n$ to $Z_{n-1}$ twisted sector.

Analogous discussion for the states corresponding to the massless
pp-wave modes is more complicated. Consider the state $|\phi_n>
\sim \psi^-_{1/2-k/n} \tilde{\psi}^+_{-1/2-k/n} |n>$ for $k>0$.
Under the assumption that the set of states corresponding to all
the massless pp-wave modes close under the action of the symmetry
algebra, we can again conclude that $G^-_{1/2}$ and
$\tilde{G}^+_{-1/2}$ annihilates this state. If we further assume
that $\tilde{G}^-_{1/2}|\phi_n>$ is proportional to
$G^+_{-1/2}|\phi_{n-1}>$ then a discussion similar to above using
the eq(\ref{conj}) would now imply that $\gamma^2=\lambda
\sqrt{N}k/n$ in agreement with the pp-wave prediction. However the
problem is that there are two different states having the same
left and right $U(1)$ charges as those of
$\tilde{G}^-_{1/2}|\phi_n>$ and $G^+_{-1/2}|\phi_{n-1}>$.
Therefore the consideration of charges is not sufficient to
conclude that the latter two states are proportional to each
other.

We conclude by noting that the issue of higher order corrections
is an important open question. The discussion we have given above
indicates that the crucial identity that one needs to prove is
equation (\ref{conj}). This equation suggests that there is an
extension of the ${\cal N}=(4,4)$ algebra which collapses to the
usual one on the physical states. This is certainly the case on
the pp-wave side and to the first order in the perturbation also
in the boundary CFT. It is also interesting to note that equation
(\ref{conj}) plays the role of the equation of motion used in
\cite{sz} to derive an all order formula for the anomalous
dimensions in the ${\cal N}=4$ Yang-Mills theory.

\begin{flushleft}
{\large\bf Appendix} \end{flushleft}
\renewcommand{\theequation}{A.\arabic{equation}}
\renewcommand{\thesection}{A.}
\setcounter{equation}{0}

In this appendix we give some details of the computation of the
correlation functions in section...  The main object we have analyzed
is the three point function denoted by $C_1$, which is given by:
\begin{eqnarray}
C_1&=&\oint_{\infty}dt_2\oint_0 dt_1 (z_2(t_2))^{-\frac{k}{n+1}}
(z_1(t_1))^{\frac{k}{n}}\sqrt{t'_1}\nonumber\\
&&<n+1|J^+(t_2)\oint_{x} dt \sqrt{t'} G^-(t)\sigma_
{\frac{1}{2},\frac{1}{2}}(x,\bar{x}) G^-(t_1)|n>,
\label{corr0}
\end{eqnarray}
where the prime on $t$ ($t_1$) means that it is differentiated with
respect to $z$ ($z_1$), giving: \be t'=\frac{1}{t^{n-1}(t-x)(n+1)}~,
\label{jac}
\ee with similar expression for $t_1$, with $x=nt_0/(n+1)$.  As
explained before, the presence of the square roots for the contour
integrations of the supercurrents is due to the fact that they have
dimension $3/2$.  Using the scalar correlator (\ref{scalar}) and the
bosonization expressions (\ref{fermion}) and (\ref{twist}) one can
easily reproduce (\ref{corr}). After rescaling $t_1$ and $t_2$ by $x$
one is left, in the limit of large $n$, with the integral:

\begin{eqnarray} C_1=\frac{1}{(n^2)u}\oint_{\infty}dt_2\oint_0 dt_1
&&(t_2)^{-\frac{kn}{n+1}}(t_2-1)^{-\frac{k}{n+1}} (t_1)^{k}
(t_1-1)^{\frac{k}{n}}\nonumber\\
&&(\frac{t_2}{t_1})^{n-1}\frac{1}{(t_2-t_1)(t_1-1)^3},
\label{corr1}
\end{eqnarray}
It is convenient to perform first the $t_2$ integral and then the
$t_1$ integral, after expanding the integrand in power series. In this
way one gets: \be C_1= \frac{1}{n^2 u}\sum_{j=1}^{n+k-1} c(k/n)_j
c(-k/n-3)_{(j-1)},
\label{int}
\ee where the coefficients $c(\alpha)_i$ are defined to be: \be
c(\alpha)_i=\frac{\Gamma(i-\alpha)}{\Gamma(-\alpha)\Gamma(i+1)}
\label{coeff}
\ee which, for small $\alpha$ behave like $c(\alpha)_i\rightarrow
-\alpha/i$, for $i\neq 0$, whereas $c(\alpha)_0=1$.  Using the above
definition of the $c_i$'s, in particular the fact that, since $j > 0$,
$c(k/n)_j$ goes like $-1/kn$, one sees that the above sum for large
$n$ is proportional to: \be \sum_{j=1}^n \frac{k}{nj} j(j+1) \sim
\frac{1}{2} n k~ .
\label{asym}
\ee We thus see that the structure constant $c_{n,n+1}$ is of order
$k/n$.

\vskip 0.5in {\bf Acknowledgements}

This project is supported in part by EEC under TMR contracts
HPRN-CT-2000-00148, HPRN-CT-2000-00122.

\rnc{\Large}{\normalsize}


\begin{thebibliography}{00}
\addcontentsline{toc}{section}{References} \frenchspacing \small
\addtolength{\itemsep}{-4pt}


\bibitem{blau}M. Blau, J. Figueroa O'Farrill, C. Hull and G.
Papadopoulos, "Penrose limits and maximal supersymmetry", arXiv:
hep-th/0201081. M. Blau, J. Figueroa O'Farrill, C. Hull and G.
Papadopoulos, "A new maximally supersymmetric background of type IIB
superstring theory", JHEP {\bf 0201} 047 (2001), arXiv:
hep-th/0110242.

\bibitem{metsaev}R.R. Metsaev, "Type IIB Green-Schwarz superstring in
plane wave Ramond-Ramond background", Nucl. Phys {\bf B625} (2002) 70,
arXiv:hep-th/0112044.

\bibitem{mt}R.R Metsaev and A.A. Tseytlin, "Exactly solvable model of
superstring in Ramond-Ramond background" arXiv: hep-th/0202109.

\bibitem{bmn}D. Berenstein, J. Maldacena and H. Nastase, "Strings in
flat space and pp-waves from ${\cal N}=4$ Super Yang-Mills", arXiv:
hep-th/0202021.

\bibitem{etal}N. Itzhaki, I.R. Klebanov and S. Mukhi, "PP Wave
Limit and Enhanced Supersymmetry in Gauge Theories", arXiv:
hep-th/0202153.
J. Gomis ans H. Ooguri, "Penrose Limit of ${\cal
N}=1$ Gauge Theories", arXiv:hep-th/0202157.
N. Kim, A. Pankiewicz, S-J. Rey and S. Theisen,
"Superstring on PP-Wave Orbifold from Large-N Quiver Theory",
arXiv:hep-th/0203080.
T. Takayanagi and S. Terashima, "Strings on
Orbifolded PP-Waves", arXiv:hep-th/0203093.
M. Alishahia and M.M. Sheikh-Jabbari, "The PP-wave
Limit of orbifolded $AdS^5\times S^5$", arXiv:hep-th/ 0203018.
E. Floratos and A. Kehagias, "Penrose Limits of
Orbifolds and Orientifolds", arXiv:hep-th/0203134.

\bibitem{bgmnn}D. Berenstein, Edi Gava, J. Maldacena, K.S. Narain and
H. Nastase, ``Open Strings on Plane waves and their Yang-Mills
duals'', arXiv:hep-th/0203249.

\bibitem{rt}J.G. Russo and A.A. Tseytlin, ``On solvable models of
type IIB superstring in NS-NS and R-R plane wave backgrounds'',
arXiv:hep-th/0202179.

\bibitem{m}J.M. Maldacena, ``The Large N Limit of
Superconformal Field Theories and Supergravity'',
Adv.Theor.Math.Phys. 2 (1998) 231; Int. J. Theor. Phys. {\bf 38}
(1999) 1113, hep-th/9711200.

\bibitem{ms}J.M. Maldacena and A. Strominger, ``
$AdS_3$ Black Holes and a Stringy Exclusion Principle'' JHEP {\bf
9812} (1998) 005.

\bibitem{db}J. de Boer, `` Six-Dimensional Supergravity on
$S^3\times AdS_3$ and 2D Conformal Field Theory'', Nucl. Phys. {\bf
B548} (1999) 139, arXiv:hep-th/9806104. J. de Boer, ``Large-N Elliptic
Genus and AdS/CFT Correspondence'', JHEP{\bf 9905} (1999) 017,
arXiv:hep-th/9812240.

\bibitem{mms}J.M. Maldacena, G. Moore and A. Strominger,
``Counting BPS Black-Holes in Toroidal type IIB String Theory'',
arXiv:hep-th/9903163.

\bibitem{agmn} E. Gava, A.B. Hammou, J.F. Morales and
K.S. Narain, ``AdS/CFT Correspondence and D1/D5 Systems in Theories
with 16 supercharges'', arXiv:hep-th/0102043.

\bibitem{hs}Y. Hikida and Y. Sugawara, ``Superstrings on PP-wave
backgrounds and symmetric orbifolds'', arXiv:hep-th/0205200.

\bibitem{lm}O. Lunin and S.D. Mathur, ``Rotating deformations of
$AdS_3\times S^3$, the orbifold CFT and strings in the pp-wave limit'',
arXiv:hep-th/0206107.

\bibitem{gms}J. Gomis, L. Motl and A. Strominger, `` PP-Wave/${\rm
CFT}_2$ Duality'', arXiv:hep-th/0206166.

\bibitem{dmvv}R. Dijkgraaf, G. Moore, E. Verlinde, and H. Verlinde,
{\it Elliptic Genera of Symmetric Products and Second Quantized
Strings }, Comm. Math. Phys. {\bf 185} (1997) 197.

\bibitem{af}G.E. Arutyunov and S.A. Frolov, ``Four graviton
scattering amplitude from $S^N{\bf R}^8$ supersymmetric orbifold sigma
model'' arXiv:hep-th/9712061.

\bibitem{lm1} O. Lunin and S.D. Mathur, ``Three-point functions for
M(N)/S(N) orbifolds with $N=4$ supersymmetry'' Comm. Math. Phys, {\bf
227} (2002) 385, arXiv:hep-th/0103169.

\bibitem{jmr} A. Jevicki, M. Mihailescu and S. Ramgoolam, ``Gravity
from CFT on $S^N(X)$ : Symmetries and Interactions'', Nucl. Phys.  {\bf
B577} (2000) 47, arXiv:hep-th/9907144.

\bibitem{lm} F. Larsen and E. Martinec, ``U(1) Charges and Moduli in the
D1/D5 System'', JHEP {\bf 9906} (1999) 019, arXiv:hep-th/9905064.

\bibitem{dmw} J.R. David, G. Mandal and S.R. Wadia, ``D1/D5 moduli in
SCFT and Gauge Theory, and Hawking Radiation'', Nucl.Phys. {\bf
B564} (2000) 103. arXiv:hep-th/9907075. J.R. David, G. Mandal and
S.R. Wadia, ''Microscopic formulation of Black Holes in string
Theor'', arXiv:hep-th/0203048.

\bibitem{sen} Ashoke Sen, ``On the Background Independence of String Field
Theory'', Nucl. Phys, {\bf B345} (1990) 551.

\bibitem{sz} A. Santambrogio and D. Zanon, ``Exact anomalous dimensions
of ${\cal N}=4$ Yang-Mills operators with large R-charges'',
arXiv:hep-th/0206079.

\end{thebibliography}
\end{document}